\documentclass[5p,onecolumn]{elsarticle}
\usepackage[utf8]{inputenc}

\usepackage{amssymb}
\usepackage{graphicx}
\usepackage{caption}
\usepackage{subcaption}
\usepackage{tabularx}
\usepackage{hyperref}
\usepackage{cleveref}
\usepackage{url}
\usepackage{booktabs}
\usepackage{xspace}
\usepackage{rotating}
\usepackage{balance}
\usepackage{listings}

\newcommand{\ie}{i.e.,\xspace}
\newcommand{\eg}{e.g.,\xspace}

\newcommand{\etal}{et al.\xspace}

\journal{Journal of Systems and Software, licensed under CC BY-NC-ND.}
\biboptions{sort&compress}

\begin{document}

\begin{frontmatter}	

\title{Towards a Catalogue of Software Quality Metrics for Infrastructure Code}
\author[JADS]{Stefano Dalla Palma}
\author[JADS]{Dario Di Nucci}
\author[Salerno]{Fabio Palomba}
\author[JADS]{Damian A. Tamburri}

\address[JADS]{Tilburg University - Jheronimus Academy of Data Science, 's-Hertogenbosch, The Netherlands}
\address[Salerno]{University of Salerno, Fisciano, Italy}

\begin{abstract}
Infrastructure-as-code (IaC) is a practice to implement continuous deployment by allowing management and provisioning of infrastructure through the definition of machine-readable files and automation around them, rather than physical hardware configuration or interactive configuration tools.
On the one hand, although IaC represents an ever-increasing widely adopted practice nowadays, still little is known concerning how to best maintain, speedily evolve, and continuously improve the code behind the IaC practice in a measurable fashion. 
On the other hand, source code measurements are often computed and analyzed to evaluate the different quality aspects of the software developed.
However, unlike general-purpose programming languages (GPLs), IaC scripts use domain-specific languages, and metrics used for GPLs may not be applicable for IaC scripts.
This article proposes a catalogue consisting of 46 metrics to identify IaC properties focusing on Ansible, one of the most popular IaC language to date, and shows how they can be used to analyze IaC scripts.
\end{abstract}


\end{frontmatter}

\section{Introduction}
The information technology market is increasingly focused on ``need for speed'': speed in deployment, faster release-cycles, speed in recovery, and more.
This need is reflected in DevOps, a family of techniques that shorten the software development cycle and intermix software development activities with IT operations~\cite{bass2015devops, artac2017devops}. 
As part of the DevOps menu, Infrastructure-as-Code (IaC)~\cite{morris2016infrastructure} \emph{``promotes managing the knowledge and experience inside reusable scripts of infrastructure code, instead of traditionally reserving it for the manual-intensive labor of system administrators which is typically slow, time-consuming, effort-prone, and often even error-prone''}. 

While IaC represents an ever-increasing widely adopted practice~\cite{morris2016infrastructure,artac2017devops,huttermann2012infrastructure}, still little is known concerning how to maintain best, speedily evolve, and continuously improve the code behind the IaC practice. Nevertheless, it is picking up more and more traction in different domains~\cite{infotalk13474,SoldaniBTMI15,lipton2018tosca}. 

A recent survey by Guerriero~\etal~\cite{GuerrieroGTP19}, conducted with industrial practitioners and experts, raised concerns about code quality and explicitly mentioned the need for instruments to support them when developing infrastructure code. 
Among others, the authors reported insights regarding IaC-specific tools and languages that are common among practitioners.
Currently, the IaC ecosystem is characterized by a plethora of different and often overlapping (in terms of their goals) tools and languages.
Thus, it is crucial to study their adoption to identify the de-facto standard ways to write IaC.
This analysis showed that no IaC tool is used by more than 60\% of respondents, with Ansible being the second most used technology (with 52.2\% of respondents using it, below the containerization technology Docker), confirming it as the de-facto standard configuration management technology.

In this context, we believe that the definition of source code metrics able to model the quality aspects of IaC could enable DevOps engineers to maintain and evolve them during Quality Assurance activities effectively.

In this article, we propose a new catalog composed of 46 measures that identify quality IaC code properties for Ansible\footnote{\url{https://www.ansible.com/}}, and showcase how they can be used to analyze infrastructure code.
The advantages of a metrics-based quality management approach to infrastructure code are manifold, among others: 

\begin{itemize}
    \item The analysis of IaC properties can help developers understand and improve the quality of their infrastructure through incremental refactoring instead of the conventional trial-and-error approach;
    \smallskip
    \item Source code properties can be used as early indicators of faulty infrastructure scripts potentially leading to expensive infrastructure failures\footnote{\url{https://www.cloudcomputing-news.net/news/2017/oct/30/glitch-economy-counting-cost-software-failures/}};
    \smallskip
    \item Specific metrics can be defined across IaC languages to understand the mutual and combined characteristics of Infrastructure-as-Code blends instead of focusing on a single vendor-locked IaC solution.
\end{itemize}

To the best of our knowledge, this work is the first to elicit a broad set of quality metrics that describe and quantify the characteristics of infrastructure code to support DevOps engineers when maintaining and evolving it.
However, it is worth noting that our catalog of metrics is broad in scope on purpose, to allow possible further studies on the relation between the proposed metrics and the quality of infrastructure code, posing the basis for a larger evaluation of infrastructure code quality.

The paper is structured as follows: in \Cref{sec:background} we briefly describe the background information of this work.
In \Cref{sec:measures} we describe the source code measures in detail.
Finally, in \Cref{sec:conclusion}, we conclude the article and present future research directions.

\section{Background and Related Work}\label{sec:background}
In this section, we briefly describe the background information to outline the context of this article and the previous works aimed at identifying source code properties that characterize IaC.

\subsection{Ansible}

The primary enabler for IaC has been the advent of cloud computing, which has made the programmatic provisioning, configuration, and management of computational resources more common.
Subsequently, many different languages and corresponding platforms have been developed, each of which deals with a specific aspect of infrastructure management: virtual machines (\eg Cloudify, Terraform), container technologies (\eg Docker Swarm, Kubernetes), configuration management tools (\eg Chef, Ansible, Puppet). 
Among them, Ansible is gaining traction in the last years as a simple and agent-less (\ie no master node) alternative to other more complex IaC technologies such as Chef and Puppet\footnote{Stemming from \url{https://github.com/search} using as search terms 'ansible', 'puppet' and 'chef' \label{footnote1}}.

Ansible is an automation engine based on the YAML language that automates cloud provisioning, configuration management, and application deployment.
It works by connecting to nodes and pushing out scripts called \textit{Ansible modules}.
Most of which describe the state of the system.
Then, Ansible executes and removes these modules when not needed. 

However, while modules allow for the proper functioning of Ansible scripts, \textit{playbooks} make possible the orchestration of multiple slices of the infrastructure topology, with exact control on the scalability of the architecture (\eg how many machines to tackle at a time).
Playbooks are essential for configuration management and multi-machine deployment in Ansible.
They can declare configurations and orchestrate steps of any manual ordered process by launching tasks within one or more \textit{plays}.
The goal of a play is to map a group of hosts to some well-defined roles, represented by Ansible \textit{tasks}, which in sum are calls to \textit{Ansible modules}.

\begin{figure}[th]
  \centering
  \includegraphics[width=\linewidth]{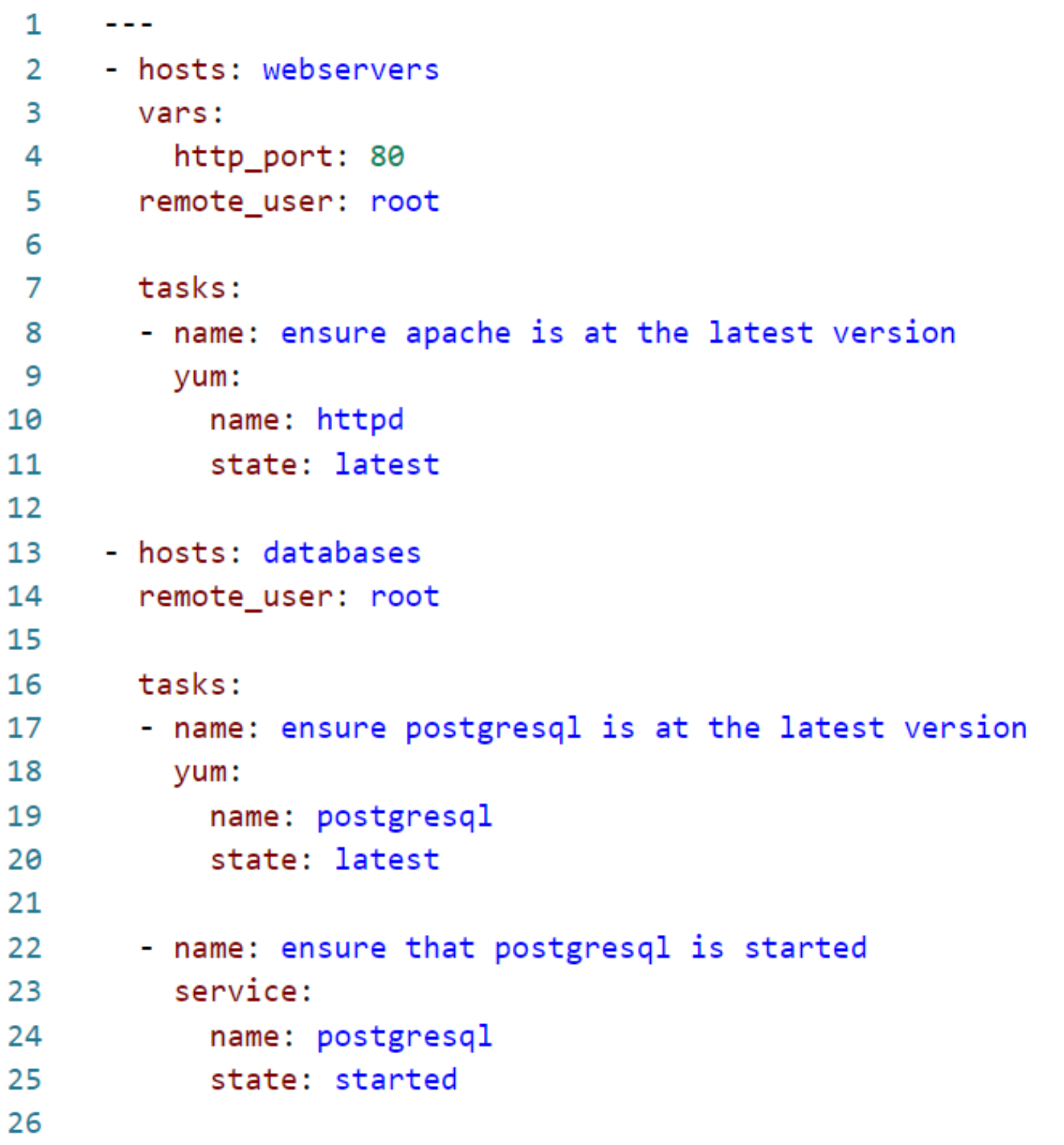}
  \caption{An example of Ansible code.}
  \label{fig:ansible-example}
\end{figure}

As an example, \Cref{fig:ansible-example} shows an Ansible code snippet representing a playbook that provisions and deploys a website\footnote{Adapted from Ansible documentation: \url{https://docs.ansible.com/ansible/latest/user_guide/playbooks_intro.html} (Accessed April 2020)}.
Hence, it configures various aspects such as the ports to be opened on the host container, the name of the user account, and the desired database to be deployed. 
It first targets the web servers to ensure that Apache server is at the latest version. Then, it checks the database servers to ensure that postgreSQL is at the latest version, and that is started.
It achieves this by mapping the hosts (lines 2 and 13) to their respective tasks (lines 8-11, 17-20, 22-25). 
There, \texttt{yum} and \texttt{service} are modules to manage packages with the yum package manager and to control services on remote hosts, respectively; 
\texttt{name} (\ie the name of the package and the database) and \texttt{state} (\ie whether present, absent, or otherwise) are parameters of these modules.
In short, by composing a playbook of multiple plays, it is possible to orchestrate multi-machine deployments and run specific commands on the machines in the \texttt{webservers} group and in those in the \texttt{databases} group.

\subsection{Related Work}
Only a few research studies have been conducted on the current development practices of infrastructure code \cite{GuerrieroGTP19} or analyzed source code properties to evaluate the quality of Infrastructure as Code \cite{spinellis-smells,RAHMAN2019148,van2018good}.
Most of the previous work describes code quality in terms of smelliness and defects-proneness of software components that are described via Chef and Puppet configuration management technologies.

Looking at code smells~\cite{folwer1999refactoring}, Sharma \etal~\cite{spinellis-smells}, Schwarz \etal~\cite{schwarz2018code}, and Rahman \etal~\cite{rahman2019seven} applied the well-know concept to IaC.
The properties identified by these studies can be grouped into: (i) \textit{Implementation Configuration} such as complex expressions and deprecated statements; (ii) \textit{Design Configuration} such as broken hierarchies and duplicate blocks~\cite{spinellis-smells}; (iii) \textit{Security Smells} such as admin by default and hard-coded secrets~\cite{rahman2019seven}; (iv) \textit{General Smells} such as long resources and too many attributes~\cite{schwarz2018code}.

As for defect prediction, Rahman \etal~\cite{RAHMAN2019148} identified ten source code measures that significantly correlate with defective infrastructure as code scripts such as properties to execute bash and/or batch commands, to manage file permissions and SSH keys, to execute external scripts.

As for IaC quality in general, Bent~\etal~\cite{van2018good} explored the notion of code quality for Puppet code by performing a survey among Puppet developers and developed a measurement model for the maintainability aspect of Puppet code quality. 
They implemented the code quality model in a software analysis tool and validated their work by a structured interview with Puppet experts and by comparing the tool results with the quality judgments of those experts.
They showed that the measurement model provides quality judgments of Puppet code that closely match the judgments of experts,
which they deemed the model appropriate and usable in practice.

In this article, we build on this line of research to define source code measures able to further model quality aspects of IaC.
We focus on Ansible, rather than Puppet and Chef, for a two-fold reason: (i) Ansible is the most popular IaC language on GitHub to date\textsuperscript{\ref{footnote1}}; (ii) at the best of our knowledge, no attempts to analyze source code properties for Ansible has been previously made.

\section{Measuring IaC Quality}\label{sec:measures}
\begin{table*}[!ht]
\centering
\caption{\textit{Ansible metrics} -- Summary of the identified and implemented metrics that identify source code properties}
\footnotesize
\begin{tabularx}{.9\linewidth}{lXl}
\toprule
\textbf{Measure} &  \textbf{Measurement techniques} & \textbf{Scope}\\ 
\midrule

\textsc{LinesBlank} & Total empty lines & General \\

\textsc{LinesComment} &  Count statements starting with \texttt{\#} & General\\

\textsc{LinesSourceCode} &  Total lines of executable code & General\\

\textsc{NumCommands}& Count of \texttt{command}, \texttt{expect}, \texttt{psexec}, \texttt{raw}, \texttt{script}, \texttt{shell}, and \texttt{telnet} syntax occurrences & General \\

\textsc{NumConditions} & Count of \texttt{is}, \texttt{in}, \texttt{==}, \texttt{!=}, \texttt{>}, \texttt{>=}, \texttt{<}, \texttt{<=} occurrences in \texttt{when} & General \\

\textsc{NumDecisions}& Count of \texttt{and}, \texttt{or}, \texttt{not} syntax occurrences in \texttt{when} & General \\

\textsc{NumDeprecatedKeywords}  & Count the occurrences of deprecated keywords & General  \\

\textsc{NumEnsure} & Count of "\texttt{\textbackslash w+.stat.\textbackslash w+ is defined}" regex matches in \texttt{when} & General \\

\textsc{NumFile} & Count of \texttt{file} syntax occurrences & General \\

\textsc{NumFileMode}  & Count of \texttt{mode} syntax occurrences & General \\

\textsc{NumLoops} & Count of \texttt{loop} and \texttt{with\_*} syntax occurrences & General \\ 

\textsc{NumMathOperations}  & Count of \texttt{+}, \texttt{-}, \texttt{/}, \texttt{//}, \texttt{\%}, \texttt{*}, \texttt{**} syntax occurrences & General \\ 

\textsc{NumParameters} & Count the keys of the dictionary representing a module & General \\

\textsc{NumPaths}  & Count of \texttt{paths}, \texttt{src} and \texttt{dest} syntax occurrences & General \\

\textsc{NumRegex} & Count of \texttt{regexp} syntax occurrences & General  \\

\textsc{NumSSH} & Count of \texttt{ssh\_authorized\_key} syntax occurrences & General \\

\textsc{NumSuspiciousComments} & Count comments with \texttt{TODO}, \texttt{FIXME}, \texttt{HACK}, \texttt{XXX}, \texttt{CHECKME}, \texttt{DOCME}, \texttt{TESTME}, or \texttt{PENDING} & General\\

\textsc{NumURLs}  & Count of \texttt{url} syntax occurrences & General \\

\textsc{NumTokens} & Count the words separated by a blank space & General \\

\textsc{NumUserInteractions} & Count of \texttt{prompt} syntax occurrences & General \\

\textsc{NumVariables } & Sum len(\texttt{vars}) in plays & General \\

\textsc{TextEntropy} & $ -\sum\nolimits_{t \in tokens(script)}{p(t) log_{2}(p(t))} $ & General \\

\textsc{NumPlays} & Count of \texttt{hosts} syntax occurrences & Playbook \\

\textsc{NumRoles} & Length of the \texttt{roles} section in a playbook & Playbook \\

\textsc{AvgPlaySize} & \textsc{LinesSourceCode}(playbook)/\textsc{NumPlays} & Playbook \texttt{|} Tasks list \\

\textsc{AvgTaskSize} & \textsc{LinesSourceCode}(tasks)/\textsc{NumTasks} & Playbook \texttt{|} Tasks list \\

\textsc{NumBlocks} & Count of \texttt{block} syntax occurrences & Playbook \texttt{|} Tasks list \\

\textsc{NumBlocksErrorHandling}  &  Count of \texttt{block-rescue-always} section occurrences & Playbook \texttt{|} Tasks list \\

\textsc{NumDeprecatedModules} & Count the occurrences of deprecated modules & Playbook \texttt{|} Tasks list  \\

\textsc{NumDistinctModules}  & Count of \textit{distinct} modules maintained by the community & Playbook \texttt{|} Tasks list\\

\textsc{NumExternalModules}  & Count occurrences of modules not maintained by the community & Playbook \texttt{|} Tasks list \\

\textsc{NumFactModules} & Count occurrences of fact modules (listed in the doc) & Playbook \texttt{|} Tasks list \\

\textsc{NumFilters} & Count of \texttt{|} syntax occurrences inside \{\{*\}\} expressions & Playbook \texttt{|} Tasks list \\

\textsc{NumIgnoreErrors}  & Count of \texttt{ignore\_errors} syntax occurrences & Playbook \texttt{|} Tasks list  \\

\textsc{NumImportPlaybook} & Count of \texttt{import\_playbook} syntax occurrences & Playbook \texttt{|} Tasks list \\

\textsc{NumImportRole}  & Count of \texttt{import\_role} syntax occurrences & Playbook \texttt{|} Tasks list \\

\textsc{NumImportTasks} & Count of \texttt{import\_tasks} syntax occurrences & Playbook \texttt{|} Tasks list \\

\textsc{NumInclude} & Count of \texttt{include} syntax occurrences & Playbook \texttt{|} Tasks list \\ 

\textsc{NumIncludeRole} & Count of \texttt{include\_role} syntax occurrences & Playbook \texttt{|} Tasks list \\  

\textsc{NumIncludeTasks} & Count of \texttt{include\_tasks} syntax occurrences & Playbook \texttt{|} Tasks list \\ 

\textsc{NumIncludeVars} & Count of \texttt{include\_vars} syntax occurrences & Playbook \texttt{|} Tasks list \\  

\textsc{NumKeys}  & Count of \textit{keys} in the dictionary representing a playbook or tasks & Playbook \texttt{|} Tasks list \\

\textsc{NumLookups} & Count of \textbf{lookup(}*\textbf{)} occurrences & Playbook \texttt{|} Tasks list  \\

\textsc{NumNameWithVariables}  & Count of \texttt{name} occurrences matching the \texttt{".*\{\{\textbackslash w+\}\}.*"} regex & Playbook \texttt{|} Tasks list\\

\textsc{NumTasks} & len(\texttt{tasks}) in playbook or tasks file & Playbook \texttt{|} Tasks list \\

\textsc{NumUniqueNames} & Count of \texttt{name} syntax occurrences with unique values & Playbook \texttt{|} Tasks list \\

\bottomrule
\end{tabularx}
\label{tab:measures}
\end{table*}

Software measurement is a quantified attribute of a characteristic of a software product~\cite{fenton2014software}.
One method to perform software measurement is to use a set of metrics to analyze the code itself.
Examples of metrics are the number of lines and functions in a single file and the number of files in an application.

\subsection{Methodology}
To extract the metrics, we applied the following methodology.
Given the novelty of the topic, we first looked for traditional and language-agnostic source code metrics that are potentially applicable to IaC.
We stemmed from a survey of almost 300 traditional and object-oriented source code metrics~\cite{nunez2017source} such as executable, commented and blank lines of code, function count, class entropy complexity, and average method size).
Needless to say that most of the object-oriented metrics do not apply to IaC: only 8 of them are present in our catalog.
Then, some of the metrics applicable to IaC were introduced by Rahman \etal~\cite{RAHMAN2019148} for Puppet and ported to Ansible. 
Finally, we searched for metrics that are specifically inherent to IaC scripts written in Ansible, starting from the \textit{atomic} structural characteristics described in the documentation\footnote{Ansible documentation \footnote{\url{https://docs.ansible.com/ansible/latest/index.html} (Accessed April 2020)}} for which structural metrics were directly implementable. 
We moved towards the more complex ones that spread through multiple scripts and/or can be expressed as a combination of atomic measures.
Those cover most of the constructs related to playbooks and tasks, such as plays, tasks, and modules, and include metrics dealing with error handling, bad and best practices (\eg using deprecated statements and naming tasks uniquely, respectively), access of data from outside sources, and more.

The search process was performed by the first author of this paper, who scanned each resource to elicit an initial set of metrics that were possibly suitable for measuring IaC code quality. 
In this step, the author considered two selection criteria: (i) the metric must be elicitable from the source analyzed, \eg when reading the Ansible documentation, the first author ensured that the description was accurate enough to enable the definition of a metric; (ii) the meaning of the metric must be clear to enable discussions on its potential impact on infrastructure code quality.
Afterward, an open discussion and card-sorting exercise \cite{LewisH10} with the remaining authors was enacted to refine the catalog and, whenever needed, assigning self-explainable names or scopes to the metrics.

More specifically, after the search, the authors opened a joint discussion on the resources retrieved.
They went through the list and analyzed the metrics to establish a rationale and possible impact that each metric may have on the quality of infrastructure code.
At this stage, they also assigned a name to the metrics: if a metric came from existing papers which already named it, they kept the same name; otherwise, they assigned a new, self-explainable name.
Since this was an open discussion, cases of disagreement related to names or rationale of the metrics were immediately discussed and solved.
In the second round, the authors proceeded by grouping metrics and discussing which of them may be considered as language-agnostic.
Also in this case, the synchronous discussion allowed them to solve cases of disagreements immediately.

At the end of the process, it was possible to classify the initial set of metrics in three groups: (i) object-oriented metrics that can be ported to Ansible, and IaC in general; (ii) metrics that were investigated in previous works on IaC and that can be ported to Ansible, and therefore to similar languages; and (iii) metrics related to best and bad practices in Ansible (which are often reported in the documentation or external resources as books).
The first two sets concern metrics that were investigated with respect to their value to code quality (even though some of them not yet studied in the context of infrastructure code); the latter set emerged when analyzing the recommendations to design quality infrastructure code.

\subsection{A Catalogue of Metrics for IaC}
Our catalog is composed of 46 code metrics. In particular, (i) 8 metrics are related to language-agnostic code characteristics, (ii) 14 metrics have been adapted by those previously developed by Rahman \etal~\cite{RAHMAN2019148} for Puppet, (iii) 24 metrics concerns some inherent characteristics of Ansible that are observable in other orchestration configuration languages as well.
The metrics related to the first group are discussed in the following.
They all concern the characterization of long/complex infrastructure code.
As widely reported in the literature on source code quality~\cite{d2012evaluating,zhang2009investigation}, those metrics could potentially make the code more prone to be defective.
While no empirical evaluation of the impact of these metrics is still available in the context of IaC, we hypothesize that similar conclusions could be reached.

\begin{itemize}
    \item \textsc{LinesSourceCode}, \textsc{LinesComment}, and \textsc{LinesBlank} to count the total number of \textit{executable lines of code}, \textit{lines of comments}, and \textit{blank lines}, respectively.
    The example in \Cref{fig:ansible-example} has 20 lines of code, no comments, and four blank lines. 
    \textit{Interpretation:} the more the lines of code, the more complex and the more challenging to maintain the blueprint.
    
    \item \textsc{NumConditions} and \textsc{NumDecisions} where a \textit{condition} is a Boolean expressions containing no Boolean operators (\eg \texttt{and} and \texttt{or}) and a \textit{decision} a Boolean expression composed of conditions and one or more Boolean operators.
    The example in \Cref{fig:ansible-example} has neither conditions nor decisions.
    In Ansible, those are mostly specified through the \texttt{when} statement that is not present in this case.
    \textit{Interpretation:} the more the conditions and decisions, the more complex and the more challenging to maintain the blueprint.
    
    \item \textsc{TextEntropy} to measure the complexity of a script based on its information content, analogous to the \textit{class entropy complexity}.
    \textit{Interpretation:} the higher the entropy, the more challenging to maintain the blueprint.
        
    \item \textsc{NumTasks} to measure the number of functions in a script, analogous to the traditional \textsc{Number of Methods Call}~\cite{nunez2017source}. We consider an Ansible  \textit{task} equivalent to a method, as its goal is to execute a module with very specific arguments. The example in \Cref{fig:ansible-example} has three tasks, so \textsc{NumTasks}=3. 
    \textit{Interpretation:} the higher the number of tasks, the more complex and the more challenging to maintain the blueprint.
    
    \item \textsc{AvgTaskSize} analogous to the \textsc{Average Method Complexity}~\cite{nunez2017source}, to measure the average size of program modules. The example in \Cref{fig:ansible-example} has three tasks. Therefore \textsc{AvgTaskSize}=4 (rounded to the nearest unit). 
    \textit{Interpretation:} the higher the more complex and the more challenging to maintain the blueprint.

\end{itemize}

Follows the second group of metrics that we generalized from the previous work conducted by Rahman \etal~\cite{RAHMAN2019148}, where the authors observed a significant correlation with defective infrastructure as code scripts.
Specifically, they conducted a qualitative analysis with practitioners and empirically validated such metrics in the scope of defect prediction of Puppet code, and a common interpretation of the following metrics is that the higher their value, the more prone to defects the blueprint:
\begin{itemize}
    
    \item \textsc{NumCommands} -- Puppet allows developers to execute external commands via the resource type \texttt{exec}. For the same functionality, Ansible provides several modules: \texttt{command}, \texttt{expect}, \texttt{psexec}, \texttt{raw}, \texttt{script}, \texttt{shell}, and \texttt{telnet}. 
    \textit{Interpretation:} the number of external commands makes the blueprint more complex and more challenging to maintain.
    
    \item \textsc{NumEnsure} -- \texttt{ensure} is a Puppet source code property used to check the existence of a file, directory or symbolic links. 
    In Ansible, the existence of those entities can be checked through the module \texttt{stat}.
    \textit{Interpretation:} blueprints with many file existence checks are less prone to misbehaviour but more challenging to test.
    
    \item \textsc{NumFile} -- \texttt{file} is a source code property used to manage files, directories, and symbolic links.
    It exists either in Puppet (as a resource type) and Ansible (as a module).
    \textit{Interpretation:} the higher this metric, the more challenging to maintain and test the blueprint.
    
    \item \textsc{NumFileMode} -- \texttt{mode} is a source code property used to set permissions of files. It exists either in Puppet (as an attribute of the \texttt{file} resource type) and Ansible (as a parameter of the \texttt{file} module).
    \textit{Interpretation:} higher levels of this metric make the blueprint less prone to misbehaviour but more challenging to test.
    
    \item \textsc{NumInclude} -- In Puppet, other scripts can be executed with the \texttt{include} function.
    This functionality in Ansible is provided by several \texttt{include} and \texttt{import} modules that allow users to break up large playbooks into smaller files, which can be used across multiple playbooks or even multiple times within the same playbook.
    \texttt{Import} statements are pre-processed at compilation-time:
    \begin{itemize}
        \item \textsc{NumImportPlaybook} -- \texttt{import\_playbook} is used to include a file with a list of plays to be executed in the current playbook.
        
        \item \textsc{NumImportRole} -- \texttt{import\_role} is used to load a \textit{role} when the playbook is parsed.
        
        \item \textsc{NumImportTasks} -- \texttt{import\_tasks} is used to import a list of tasks to be added to the current playbook for subsequent execution.
    \end{itemize}
    
    \texttt{Include} statements are processed at execution-time\footnote{\url{https://docs.ansible.com/ansible/latest/user_guide/playbooks_reuse_includes.html}}:

    \begin{itemize}
        \item \textsc{NumInclude} --\texttt{include} is used to include a file with a list of plays or tasks to be executed in the current playbook.
        
        \item \textsc{NumIncludeRole} -- \texttt{include\_role} is used to dynamically load and execute a specified role as a task.
        
        \item \textsc{NumIncludeTasks} -- \texttt{include\_task} is used to include a file with a list of tasks to be executed in the current playbook.
        
        \item \textsc{NumIncludeVars} -- \texttt{include\_vars} is used to load YAML or JSON variables from a file or directory, recursively, during task run-time. 
    \end{itemize}
    \textit{Interpretation:} on the one hand, the more the includes and imports, the higher the reusability. On the other hand, the more the includes and imports, the more difficult to maintain the blueprint.

    \item \textsc{NumParameters} -- In Puppet, the state of a resource is described with an attribute.
    Similarly, in Ansible \textit{parameters} (or arguments), describe the desired state of the system.
    \textit{Interpretation:} the more the parameters, the more challenging to debug and test the blueprint.
    
    \item \textsc{NumSSH} -- \texttt{ssh\_authorized\_key} is a Puppet source code property used to manage SSH authorized keys.
    The analog in Ansible is the module \texttt{authori\-zed\_key}, used to add or remove SSH authorized keys for particular user accounts. \textit{Interpretation:} the more this property, the more challenging to maintain and test the blueprint.
    
    \item \textsc{NumURLs} -- URL refers to URLs used to specify a configuration.
    Ansible defines a module called \texttt{uri} to interact with \textit{http} and \textit{https} web services, and requires to set a parameter \texttt{url}.
    \textit{Interpretation:} the more the URLs, the more challenging to maintain and test the blueprint.

\end{itemize}

\begin{figure}[ht]
    \centering
    \includegraphics[width=\linewidth]{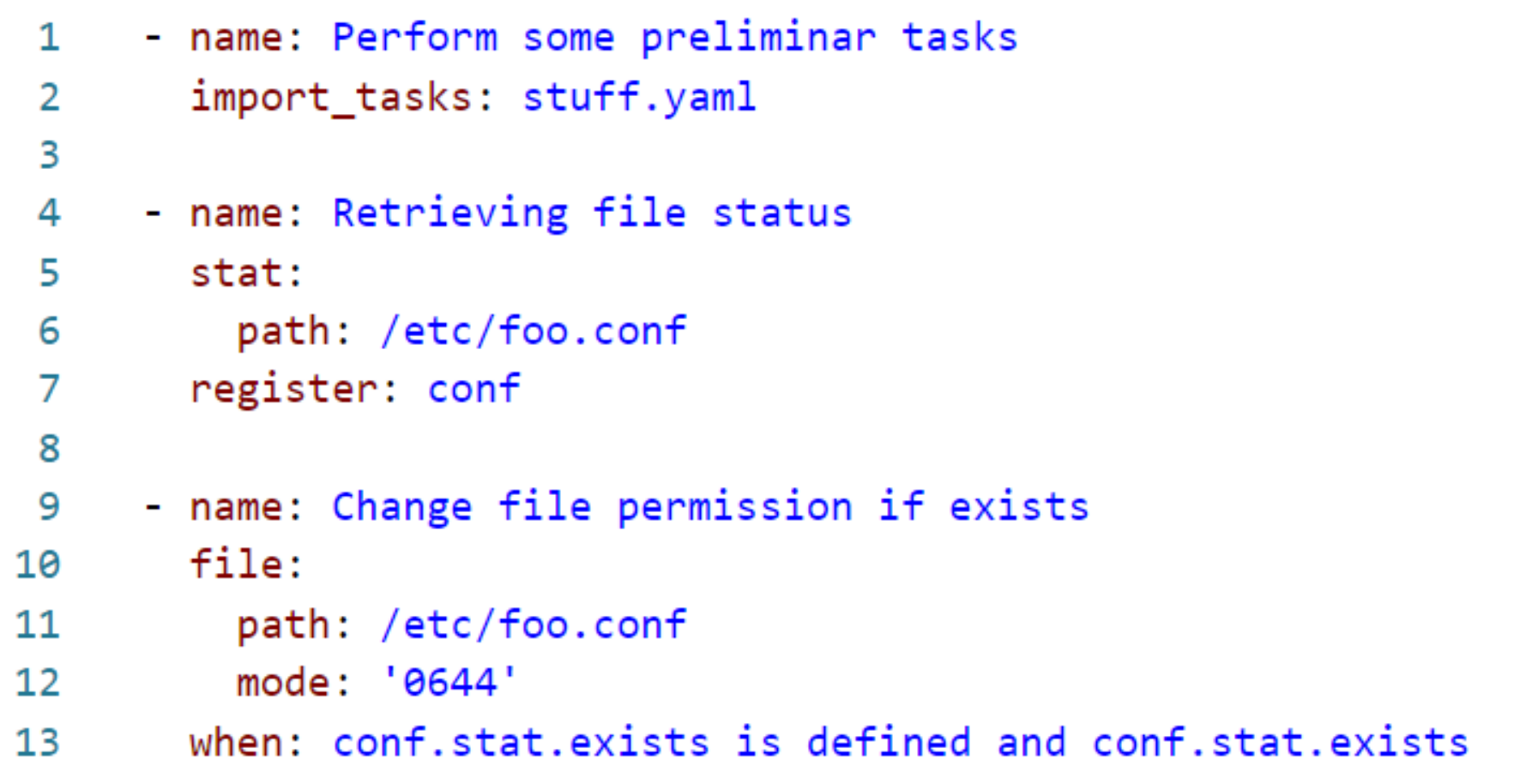}
    \caption{Some of the source code properties designed by Rahman \etal~\cite{RAHMAN2019148} for Puppet and ported to Ansible}
    \label{fig:metrics_from_puppet}
\end{figure}

\Cref{fig:metrics_from_puppet} shows a code snippet on which we computed such metrics.
In this example, \textsc{NumInclude} = 1 (line 2), \textsc{NumParameters} = 3 (lines 6, 11, 12), \textsc{NumFile} = 1 (line 10), \textsc{NumFileMode} = 1 (line 12), and \textsc{NumEnsure} = 1 (line 13); \textsc{NumSSH} = 0; \textsc{NumURLs} = 0.
It also has two conditions bound by a logic \texttt{and} at line 11, consequently \textsc{NumConditions} = 2 and \textsc{NumDecisons} = 1.

\medskip

\begin{figure}[hbt]
    \centering
    \includegraphics[width=\linewidth]{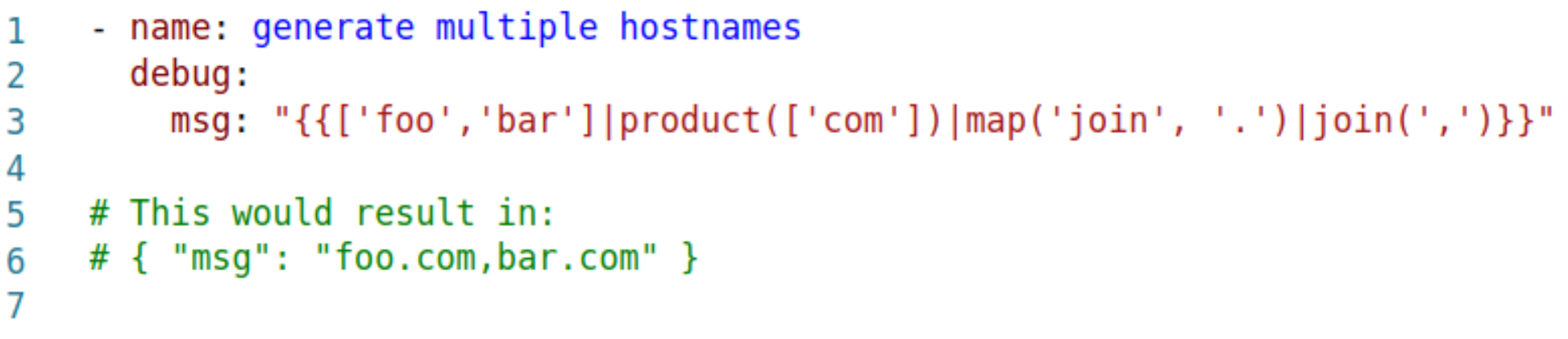}
    \caption{The product filter returns the cartesian product of the input iterables, that is roughly equivalent to nested for-loops.}
    \label{fig:ansible_filters}
\end{figure}

The third group of metrics was derived from the Ansible documentation and are mainly related to best and bad practices and (external) data management.
These following metrics could affect the quality of infrastructure code in terms of comprehensibility and maintainability:

\begin{itemize}

   \item \textsc{DeprecatedKeywords} and \textsc{DeprecatedModu\-les} -- Deprecated modules and keywords usage is discouraged as they are kept for backward compatibility only.
   \textit{Interpretation:} the more the deprecated keywords and modules, the more difficult it is to maintain and evolve the code. In addition, the higher the likelihood to crash if the current system does not support retro-compatibility.
    
   \item \textsc{NumBlocks} and \textsc{NumBlocksErrorHandling} -- A \textsc{block} logically groups tasks within a section, but also allows for exception handling by appending a \texttt{rescue} or an \texttt{always} to the block. 
   The tasks in the block are typically executed.
   A \texttt{rescue} section is executed only when an error is raised, while an \texttt{always} section is executed in any case (\ie included in case of errors). \textit{Interpretation:} the more the blocks, the easier code maintenance. 
   However, many blocks without rescue/always sections increase the system's chance to crash when incurring in wrong behaviours.
    
    \item \textsc{NumDistinctModules} -- Modules are reusable and standalone scripts called by tasks. They allow to change or get information about the state of the system and can be interpreted as a degree of responsibility of the blueprint.
    Therefore, we hypothesize that a blueprint consisting of many distinct modules is less self-contained and potentially affect the complexity and maintainability of the system, as it is responsible for executing many different tasks rather than a task several times with different options, for example, to ensure the presence of dependencies in the system (as the \texttt{yum} module in \Cref{fig:ansible-example}).\\
    \textit{Interpretation:} the more the distinct modules, the more challenging to maintain the blueprint.
    
    \item \textsc{NumExternalModules} -- Many modules are maintained by the community. However, users can create and maintain their modules, called \textit{external modules}. While modules maintained by the community require to be fully documented and tested, this is not needed for external modules. Therefore, we conjecture that a blueprint with a high number of external modules is more challenging to maintain than a blueprint containing only modules maintained by the community. \textit{Interpretation:} The more the external modules, the more challenging to maintain the blueprint and the higher the chance of system's misbehaviour.
    
    \item \textsc{NumFactModules} -- \textit{Fact modules} are modules that do not change the state of the system but only return data. We hypothesize that blueprints with many fact modules are less prone to unexpected behaviours and easier to test, as they do not alter the system's state. \textit{Interpretation}: the more the \texttt{fact} modules the easier to test and maintain the blueprint.
    
    \item \textsc{NumFilters} -- Filters transform the data of a template expression, for example, for formatting or rendering them. Although they allow for transforming data in a very compact way, filters can be concatenated to perform a sequence of different actions, as shown in \Cref{fig:ansible_filters}. We believe that this aspect may potentially affect the readability and maintainability of the code. \textit{Interpretation:} the more the filters, the lower the readability and the more challenging to maintain the blueprint.
    
    \item \textsc{NumIgnoreErrors} -- Ansible provides different ways to handle errors.
    Among others, it is possible to prevent a playbook from stopping when a task fails by setting \texttt{ignore\_errors:True}.
    However, ignoring errors is considered as a bad practice, since it hides error handling. 
    \textit{Interpretation:} the more the \texttt{ignore\_errors:True} statements, the more the system is hard to debug and prone to misbehaviour.
    
    \item \textsc{NumLookups} -- Lookups are an advanced feature that allows access to outside data sources and requires an advanced working knowledge of Ansible plays before incorporating them. Some lookups pass arguments to a shell, and one should use them carefully to ensure safe usage. \textit{Interpretation:} the more the lookups, the higher the chance of system's misbehaviour and the more challenging to maintain.
    
    \item \textsc{NumSuspiciousComments} -- Suspicious comments warn the presence of defects, missing functionality, or the system's weakness. \textit{Interpretation:} the more the suspicious comments, the higher the chance of system's misbehaviour because of missing or incomplete functionalities.
    
    \item \textsc{NumUniqueName} -- Uniquely naming plays and tasks is a best practice to locate problematic tasks quickly.
    Duplicate names may lead to not deterministic or at least not obvious behaviours ~\cite{keating2015mastering}. \textit{Interpretation:} the more the entities with unique names the higher the maintainability and readability of the blueprint.
    
    \item  \textsc{NumNamesWithVariables} -- Having uniqueness as a goal, many playbook developers prefer to use variables instead of hard-coding names. This strategy may work well, but authors need to take care of the source of the variable data they are referencing. Indeed, variable data can come from various locations, and the values assigned to variables can be defined many times. For the sake of play and task names, only variables for which the values can be determined at playbook parse time will parse and render correctly. If the data of a referenced variable is discovered via a task, the variable string will be displayed unparsed in the output~\cite{keating2015mastering}, potentially affecting debugging and software auditing. \textit{Interpretation:} although names with variable could make more succinct code, they could hinder code debugging and potentially lead to system's misbehaviour.
     
    \item \textsc{NumUserInteractions} -- In some cases, an Ansible script requires the user input (\eg username and password to access a service). 
    Asking for external input may potentially affect the correctness of the program at run-time. User interactions have to be handled by the program with several conditions. We conjecture that, if not handled properly, a given input may lead the system to crash at run-time.\\ \textit{Interpretation:} the more user interactions, the higher the chance of system's misbehaviour.

\end{itemize}

The remaining metrics are self-describing and measure different aspects of the size of a blueprint which may affect its quality in terms of complexity and readability: \textsc{NumPlays}, \textsc{AvgPlaySize}, \textsc{NumRoles}, \textsc{NumVariables}, \textsc{NumLoops}, \textsc{NumMathOperations}, \textsc{NumPaths}, \textsc{NumRegex} (\ie regular expressions), \textsc{NumTokens} (\ie words separated by blank spaces), and \textsc{NumKeys} (\ie keys of the dictionary representing a playbook or a list of tasks).
In particular, paths and regular expressions are often subject to typos, which might lead to run-time errors if they are not correctly handled. We conjecture that the more they are, the higher the chance the system will run into unexpected behaviour.
In general, \textit{we hypothesize that the higher the number of the aforementioned source code properties, the more complex the blueprint.}

The complete list of the metrics we designed is shown in \Cref{tab:measures}.
The code metrics can be categorized in terms of their scope (\ie \textit{general}, \textit{playbook}, and \textit{tasks list}) depending on the particular construct and/or artifact they target.
\textit{General} metrics can apply to either playbooks and task files, \ie files that contain a flat list of tasks, but can also generalize to other languages.
\textit{Playbook} metrics operate within a single playbook; while \textit{tasks list} metrics can operate either within a playbook (when the construct \textit{tasks} is present) and within tasks files.

\begin{figure}[ht]
    \centering
    \includegraphics[width=\linewidth]{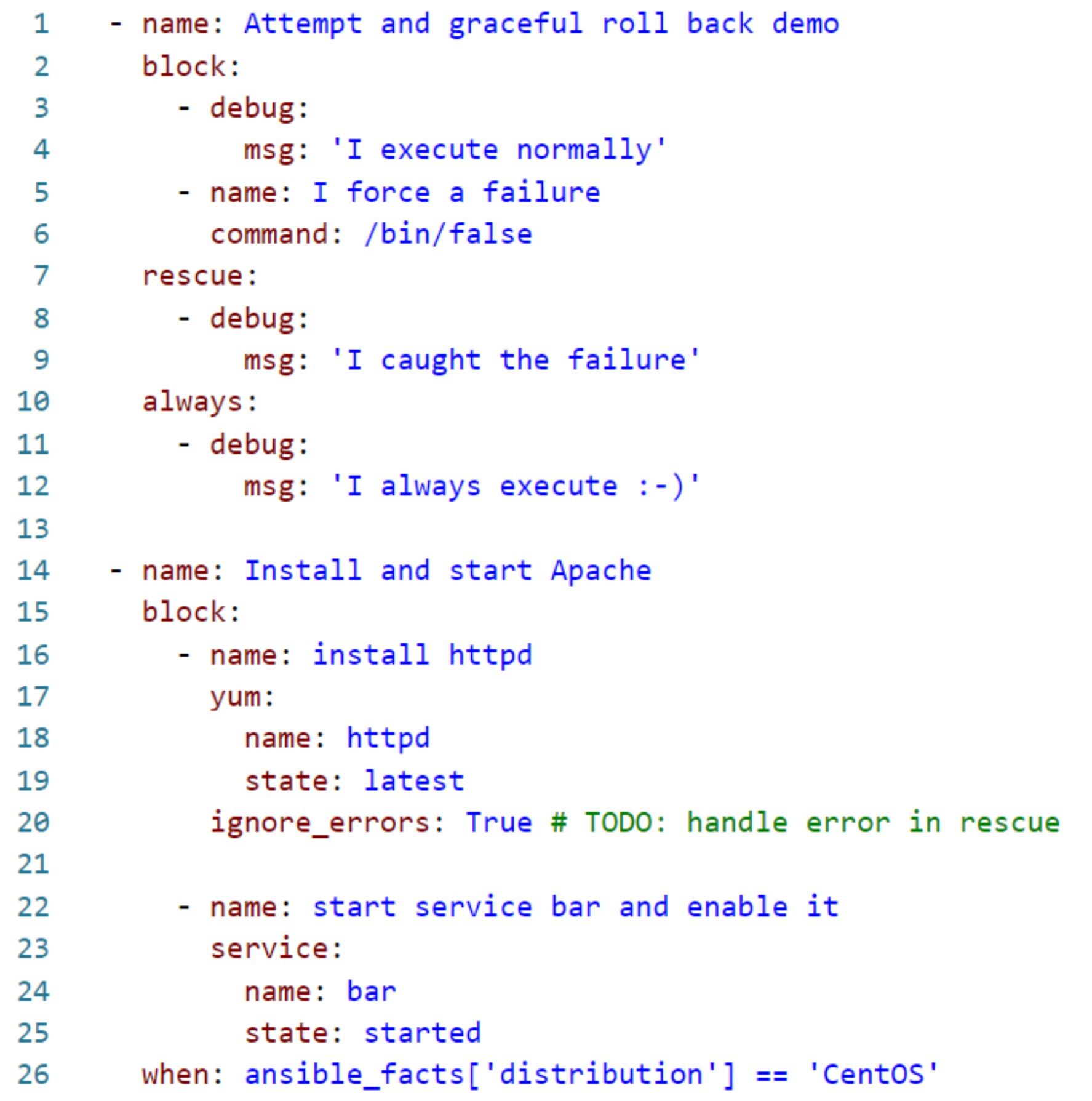}
    \caption{An example of properties inherent to Ansible.}
    \label{fig:ansible_example_2}
\end{figure}

\Cref{fig:ansible_example_2} shows an Ansible code snippet that consists of a list of tasks adapted from the Ansible documentation\footnote{\url{https://docs.ansible.com/ansible/latest/user_guide/playbooks_blocks.html (Accessed on April 2020)}}. 
The tasks are grouped in two blocks: the first block (lines 3-8) is used to handle errors along with the \texttt{rescue} and \texttt{always} sections (lines 8 and 11); the second block (lines 16-27) is used to allow the tasks' execution only if the current system distribution is 'CentosOS' (line 27). Therefore, \textsc{NumBlocks} = 2 and \textsc{NumBlocksErrorHandling} = 1.
The \texttt{ignore\_errors:True} at line 21 prevents a playbook from stopping if something goes wrong with the Apache server's installation. 
The error should not be ignored but caught and handled in a \texttt{rescue} section.
The word \texttt{TODO} in the comment is suspicious and suggests a possible issue.
Here \textsc{NumIgnoreErrors} = 1 and \textsc{NumSuspiciousComments} = 1.
Other information that can be extracted from the snippet relate to the number of distinct modules, namely \texttt{debug}, \texttt{command}, \texttt{yum} and \texttt{service}, and the the number of unique names (lines 2, 6, 15, 17, 23). 
Note that the \texttt{name} at lines 19 and 25 are not tasks' names, but module parameters.
The metrics calculated here are \textsc{NumDistinctModules} = 4 and \textsc{NumUniqueNames} = 5.

\section{Conclusion and Research Roadmap}
Infrastructure code is emerging as the de-facto challenge for software maintenance and evolution of the coming years, especially concerning complex systems blending microservices with serverless components.
Indeed, varied and elaborate infrastructure designs may well change dynamically during operations.
On the one hand, infrastructure source code properties may be used and combined as surrogate metrics for defect-proneness of infrastructure components and to identify smells and refactoring operations during Quality Assurance activities connected to infrastructure code. 
On the other hand, the existing code measures cannot currently and directly model the aforementioned aspects of IaC.

We propose a broad catalogue of 46 structural-based code measures to evaluate the different aspects of IaC, the most comprehensive measures set for IaC to the best of our knowledge. Although our implementation targets particularly Ansible, the aforementioned measures are equivalent (and portable) to other languages (\eg Chef, Puppet) and offer a general-purpose metrics-based approach for IaC quality evaluation.
Given the early-stage research on IaC analytics, there are several avenues for future work.

\paragraph{Empirically Investigating the Relationship between IaC  Metrics and Code Quality}
We are aware that some of the proposed metrics may have little effect on code quality. To accurately identify and validate what metrics significantly affect a given quality aspect of a configuration management system, future empirical studies are required.
First, more research is needed to understand the relation between the presented metrics and the quality of IaC blueprints. 
Second, there is still a lack of empirical evidence regarding the scope and usability of such measures, \eg whether they can be used and/or combined to detect code smells and bugs. A step in this direction is the evaluation of the proposed metrics as a proxy of a defect prediction model for IaC scripts.
A further step would be the mapping of the proposed metrics to the software quality attributes they actually model (\eg maintainability, complexity, reusability).

\paragraph{IaC Metrics in Software Defect Prediction}
Software Defect Prediction~\cite{hall2011systematic} helps to identify the parts of the systems that may require more testing because prone to defects.
The definition of effective prediction of defect-prone blueprints in the Infrastructure-as-Code scope may help DevOps engineers focus on such demanding blueprints during testing and maintenance activities, thus allocating effort and resources more efficiently.
Product and process metrics (\ie metrics that capture aspects concerning the development process rather than the code itself, for example the number of modification to a file) have been widely adopted in defect prediction of systems written in General Purpose Languages, and in many cases process metrics over-performed product metrics in terms of classification performance~\cite{moser2008comparative, rahman2013and}. Nevertheless, little is known about the prediction power of both sets of metrics across domain-specific languages such as IaC. 
As opposed to product metrics, process metrics are language-agnostic and do not need to be ported to IaC (\ie the original ones can be used). Therefore, the definition of the proposed catalog would allow for a fair comparison between traditional process metrics and IaC-oriented product metrics in the scope of defect prediction of infrastructure code.

\paragraph{IaC Metrics Generalization}
Afterwards, we plan to generalise our catalogue to support other configuration orchestration languages.
To this aim, we will map the Ansible-specific characteristics to other languages.
Among the others, we plan to support the de-iure industry standard for infrastructure code, namely, the "Topology and Orchestration Specification for Cloud Applications" TOSCA~\cite{tosca2019} language, a YAML-based OASIS standard for defining infrastructure topologies.
This specification was initially designed as an open standard for formatting templates. So, tasks (\eg cloud resource deployment and orchestration) could be translated into a generally readable form and become more portable across platforms.
Overall, the standard aims to make it easier to update, extend, or move cloud-based resources, thus opening up opportunities for building a universal software-quality model for configuration orchestration languages.\label{sec:conclusion}

\section*{Acknowledgements}
The authors would like to thank the colleagues and anonymous reviewers for their invaluable feedback.
Stefano, Dario, and Damian are supported by the European Commission grant no. 825040 (H2020 RADON).
Fabio acknowledges the support of the Swiss National Science Foundation through the SNF Project No. PZ00P2\_186090 (TED). 

\balance

\bibliographystyle{elsarticle-num}
\bibliography{main}
\end{document}